\documentclass[twocolumn,showpacs,preprintnumbers,amsmath,amssymb,APSl,prd,nofootinbib,superscriptaddress]{revtex4-1}

\usepackage{dcolumn}
\usepackage{bm}
\usepackage{ifpdf}
\usepackage{hyperref}
\usepackage{dcolumn}
\usepackage{bm}
\usepackage[spanish,english]{babel}
\usepackage{amsfonts}
\usepackage{amssymb}
\usepackage{graphicx}
\usepackage{color}
\usepackage[latin1]{inputenc}

\newcommand{\be}{\begin{equation}}
\newcommand{\en}{\end{equation}}
\newcommand{\bea}{\begin{eqnarray}}
\newcommand{\ena}{\end{eqnarray}}
\newcommand{\bes}{\begin{subequations}}
\newcommand{\ees}{\end{subequations}}

\definecolor{red}{rgb}{1,0,0}

\def\p{\partial}

\def\+{^\dagger}

\def\<{\leftarrow}
\def\>{\rightarrow}

\def\({\left(}
\def\){\right)}

\def\a{\alpha} \def\b{\beta}  \def\d{\delta} \def\e{\epsilon}
\def\m{\mu} \def\n{\nu}   \def\l{\lambda} 
\def\k{\kappa}\def\G{\Gamma}

\def\q{\quad}

\def\W{{\Omega}}
\newcommand{\LL}{{\cal L}}

\newcommand{\bi}{\begin{itemize}} 				\newcommand{\ei}{\end{itemize}}
\newcommand{\benu}{\begin{enumerate}} 		\newcommand{\enu}{\end{enumerate}}
\newcommand{\bd}{\begin{dinglist}{0}}     \newcommand{\ed}{\end{dinglist}}
\newcommand{\bfig}{\begin{figure}[htbp]}  \newcommand{\efig}{\end{figure}}
        			
\newcommand{\bc}{\begin{center}} 				  \newcommand{\ec}{\end{center}}
\newcommand{\bsub}{\begin{subequations}}  \newcommand{\esub}{\end{subequations}}
\newcommand{\ben}{\begin{eqnarray}} 			\newcommand{\een}{\end{eqnarray}}
\newcommand{\ba}[1]{\begin{array}{#1}} 		\newcommand{\ea}{\end{array}}
\newcommand{\eea}{\end{array}\end{equation}}

\begin{document}
\title{Mapping Ricci-based theories of gravity into general relativity}

\author{V. I. Afonso} \email{viafonso@df.ufcg.edu.br}
\affiliation{Unidade Acad\^{e}mica de F\'isica, Universidade Federal de Campina
Grande, 58109-970 Campina Grande, PB, Brazil}
\affiliation{Departamento de F\'{i}sica Te\'{o}rica and IFIC, Centro Mixto Universidad de Valencia - CSIC.
Universidad de Valencia, Burjassot-46100, Valencia, Spain}
\author{Gonzalo J. Olmo} \email{gonzalo.olmo@uv.es}
\affiliation{Departamento de F\'{i}sica Te\'{o}rica and IFIC, Centro Mixto Universidad de Valencia - CSIC.
Universidad de Valencia, Burjassot-46100, Valencia, Spain}
\affiliation{Departamento de F\'isica, Universidade Federal da
Para\'\i ba, 58051-900 Jo\~ao Pessoa, Para\'\i ba, Brazil}
\author{D. Rubiera-Garcia} \email{drgarcia@fc.ul.pt}
\affiliation{Instituto de Astrof\'{\i}sica e Ci\^{e}ncias do Espa\c{c}o, Faculdade de
Ci\^encias da Universidade de Lisboa, Edif\'{\i}cio C8, Campo Grande,
P-1749-016 Lisbon, Portugal}

\date{\today}
\begin{abstract}
We show that the space of solutions of a wide family of Ricci-based metric-affine theories of gravity can be put into correspondence with the space of solutions of general relativity (GR). This allows us to use well-established  methods and results from GR to explore new gravitational physics beyond~it. \end{abstract}
\maketitle

\section{Introduction}

With the advent of gravitational wave astronomy resulting from the findings of the LIGO-Virgo Collaboration \cite{Abbott:2016blz,Abbott:2017oio,Abbott:2017nn}, a new window on the most energetic events in nature is now open. Among the fascinating discoveries yet to come, multimessenger observations may allow us to explore subtle departures from the established predictions for compact objects as well as the core principles of GR \cite{AbbottNS}. In particular, the so-called black hole mimickers, such as boson stars \cite{Liebling:2012fv}, gravastars \cite{Visser:2003ge}, traversable wormholes \cite{Cardoso:2016rao} and black stars \cite{Brito:2015yga}, among others, and which share in common the absence of an event horizon, are potential alternatives to black holes that may affect our current interpretation of certain astrophysical phenomena as well as our understanding of fundamental physics \cite{Cardoso:2017cqb,Abedi:2016hgu,Cardoso:2016oxy}. Some of these objects arise in extensions of GR and may be supported by standard matter, not requiring exotic sources for their existence. Extensions of GR are indeed motivated by multiple reasons, and a plethora of alternative gravity models currently exist in the literature \cite{Nojiri:2017ncd, Berti:2015itd,Capozziello:2011et,Clifton:2011jh, DeFelice:2010aj}. However, confronting gravitational wave data with the predictions of those theories represents a formidable challenge. In fact, the development of numerical methods and algorithms is strongly conditioned by their implementation in the framework of GR \cite{Shibata:2011jka,Rezzolla:2010fd,Pretorius:2004jg,Baiotti:2004wn,Baumgarte:2002jm,Lehner:2001wq}. Their extension to other theories is expensive in many respects and, in practice, it may preclude the use of gravitational wave data to unveil subtle departures from the predictions of GR.

In this work we present a (broad) class of gravity theories whose analysis and confrontation with observations can be carried out systematically by borrowing techniques and methods previously developed in the framework of GR. This is possible thanks to the existence of a correspondence between the space of solutions of GR and the space of solutions of those theories. We show how to obtain this correspondence, in general, for some matter sources and illustrate the process with a particular gravity theory.

\section{Theoretical framework}

We focus on a family of theories of the form
\begin{equation}\label{eq:f-action}
\mathcal{S}=\int d^4x \sqrt{-g} \mathcal{L}_G\left[g_{\mu\nu},R_{(\mu\nu)}(\Gamma)\right]+\mathcal{S}_m[g_{\mu\nu},\psi]   \ ,
\end{equation}
where $\mathcal{S}_m$ denotes the (minimally coupled) matter action, with $\psi_m$ labeling collectively the matter fields, $g$ is the determinant of the spacetime metric $g_{\mu\nu}$, $\mathcal{L}_G\left[g_{\m\n},R_{(\mu\nu)}(\G)\right]$ is the gravity Lagrangian, constructed out of the metric and the (symmetrized) Ricci tensor, $R_{\mu\nu}={R^\alpha}_{\mu\beta\nu} \delta^\beta_\alpha$, where  $ {R^\alpha}_{ \beta\mu\nu}=\p_\m\G_{\n\b}^\a-\p_\n\G_{\m\b}^\a+\G_{\m\l}^\a\G_{\n\b}^\l-\G_{\n\l}^\a\G_{\m\b}^\l$ is the Riemann tensor of a connection $\Gamma \equiv \Gamma_{\mu\nu}^{\lambda}$, which we assume is a priori independent of the metric (metric-affine or Palatini approach).
We emphasize that in order to guarantee that $\mathcal{L}_G$ is a scalar function, its functional dependence on the metric and the Ricci tensor must be through traces of powers of the object ${M^\mu}_\nu\equiv g^{\mu\alpha}R_{(\alpha\nu)}(\G)$ \cite{Jimenez:2014fla}. Within this class of theories, which will be referred to as Ricci-based gravities (RBGs, implicitly assumed hereafter to be metric-affine formulated), we find GR itself, $f(R)$ and $f(R,R_{(\mu\nu)}R^{(\mu\nu)})$ theories \cite{Olmo:2011uz}, Born-Infeld inspired models \cite{BHOR}, \ldots,  among others, all of which have received much attention in the literature.

These theories admit an Einstein-frame representation \cite{BJEf} in terms of an auxiliary metric $q_{\mu\nu}$ whose relation with the spacetime metric $g_{\mu\nu}$ can be parameterized as $q_{\mu\nu}=g_{\mu\alpha}{\Omega^\alpha}_\nu$, with the matrix ${\Omega^\alpha}_\nu$ a function of the  matter fields (and possibly of $g_{\mu\nu}$) whose specific details depend on the particular $\mathcal{L}_G$ chosen. By performing independent variations with respect to the metric and connection, it can be shown  that $\Gamma^\alpha_{\mu\nu}$ is the Levi-Civita connection of $q_{\mu\nu}$, ${M^\mu}_\nu$  turns out to be an algebraic function of ${T^\mu}_\nu$, and the field equations can finally be written in the Einstein-like form \cite{BHOR,Afonso:2017bxr}
\begin{equation}\label{eq:GmnGeneral}
{G^\mu}_\nu(q)=\frac{\kappa^2}{|\hat{\Omega}|^{1/2}}\left[{T^\mu}_\nu-{\delta^\mu}_\nu\left(\mathcal{L}_G+\tfrac{T}{2}\right)\right] \ .
\end{equation}
Here ${G^\mu}_\nu(q) \equiv  q^{\mu\alpha}R_{\alpha\nu}(q) - \frac{1}{2} {\delta^\mu}_{\nu} R(q)$, $\kappa^2\equiv 8\pi G$,
a hat denotes a matrix and vertical bars its determinant, and ${T^\mu}_\nu\equiv g^{\mu\alpha}T_{\alpha\nu}$.
It is important to note that, due to the algebraic relation that exists between ${M^\mu}_\nu$ and ${T^\mu}_\nu$,  in RBGs $\mathcal{L}_G$ and $|\hat{\Omega}|^{1/2}$ can be written on-shell as functions of ${T^\mu}_\nu$. As a result, the right-hand side of (\ref{eq:GmnGeneral}) is just a function of the matter and the metric, and the vacuum field equations boil down to those of GR (possibly with a cosmological constant). This has two remarkable consequences. First, RBGs do not propagate extra degrees of freedom beyond the two tensor (spin-2) polarizations of GR, which makes them automatically compatible with the results of the LIGO-Virgo network \cite{Abbott:2017oio}, where purely tensor modes were reported to be strongly favoured by data over purely vector or purely scalar modes. Second, gravitational waves propagate at the speed of light in vacuum, thus allowing RBGs to survive the slaughter of modified theories of gravity \cite{Baker:2017hug,Creminelli:2017sry,Ezquiaga:2017ekz} resulting from the almost simultaneous observation of GW170817 and GRB170817 \cite{AbbottNS}.

\section{Gravity with a scalar field}

Let us consider as the matter source of our theories a (complex) scalar field with $U(1)$ symmetry
described by the action $\mathcal{S}_m=-\frac{1}{2}\int d^4x \sqrt{-g} \LL_\Phi$, where $\mathcal{L}_\Phi=X_\Phi+2V(\Phi, \Phi^{\ast} )$,  $X_\Phi\equiv g^{\alpha\beta}\partial_\alpha \Phi^{\ast} \partial_\beta \Phi$ is the kinetic term, and $V$ the potential. Scalar fields are of particular interest for astrophysics, as they can support compact objects such as solitonic boson stars \cite{Liebling:2012fv} or hairy black holes \cite{Herdeiro:2015waa}, and play a prominent role in their dynamics, for instance via superradiance \cite{Brito:2015oca}, scalar clouds \cite{Sanchis-Gual:2015sxa} or bosenova explosions \cite{Sanchis-Gual:2015lje}. They are also very relevant in inflationary cosmology \cite{Baumann:2009ds, Lyth:2009zz}, dark energy models \cite{Copeland:2006wr,Peebles:2002gy}, and higher-dimensional models of the braneworld type  \cite{Randall:1999vf,Goldberger:1999uk,DeWolfe:1999cp,Maartens:2010ar}.

The energy-momentum tensor for this scalar field is given by
\begin{equation}\label{eq:TmnPhi}
{T^\mu}_\nu=g^{\mu\alpha}\Phi_{,\alpha} ^{\ast}\Phi_{,\nu}-\frac{\mathcal{L}_\Phi}{2}{\delta^\mu}_\nu \ ,
\end{equation}
where $\Phi_{,\nu}\equiv \partial_\nu \Phi$. Inserting \eqref{eq:TmnPhi} in (\ref{eq:GmnGeneral}), we get
\begin{equation}\label{eq:GmnScalar}
{G^\mu}_\nu(q)=\frac{\kappa^2}{|\hat{\Omega}|^{\frac{1}{2}}}\left[g^{\mu\alpha}\Phi^{\ast}_{,\alpha} \Phi_{,\nu}- {\delta^\mu}_\nu  \left(\mathcal{L}_G-V\right)\right] \ .
\end{equation}
Once the gravity Lagrangian $\mathcal{L}_G$ is specified, an algebraic  relation (nonlinear in general) between the matrix ${\Omega^\mu}_\nu$ and the matter ${T^\mu}_\nu$ can be obtained, which allows us to express both $\mathcal{L}_G$ and $|\hat{\Omega}|$ as functions of the matter. This nonlinear relation can be seen as an infinite power series expansion in terms of  ${T^\mu}_\nu$. Now, in the scalar field case, ${T^\mu}_\nu$ has a term proportional to the identity plus another one linear in ${X}^\mu_{\ \nu}\equiv g^{\mu\alpha}\Phi^{\ast}_{,\alpha} \Phi_{,\nu}$, whose powers are proportional to itself, namely, $\hat{{X}}^n = {X}_{\Phi}^{n-1} \hat{{X}}$. This leads to a dramatic simplification of the series expansion, which retains the same algebraic structure as ${T^\mu}_\nu$, namely,
\begin{equation}\label{eq:scalar}
{\Omega^\mu}_{\nu}=C_1[{X}_{\Phi},V]{\delta^\mu}_{\nu} +C_2[{X}_{\Phi},V]{X}^\mu_{\ \nu}   \ ,
\end{equation}
with the $C_i[{X}_{\Phi},V]$ being model-dependent functions. The determinant $|\hat{\Omega}|$ can be computed in terms of traces of powers of  ${\Omega^\mu}_{\nu}$ and takes the form $|\hat{\Omega}|= C_1^3(C_1+C_2 {X}_{\Phi})$. Once the theory is fully specified, one can proceed similarly with $\mathcal{L}_G$, which should be a function of $X_\Phi$ and $V$ as well. As a result, the right-hand side of (\ref{eq:GmnScalar}) is a highly nonlinear function of the ${T^\mu}_\nu$ of $\Phi$.

Now we will show that the right-hand side of (\ref{eq:GmnScalar}) can be written as the usual ${T^\mu}_\nu$ of another field $\phi$ coupled to $q_{\mu\nu}$, thus turning the original modified gravity problem (highly nonlinear in the ${T^\mu}_\nu$ of $\Phi$) into a standard GR problem (linear in the ${T^\mu}_\nu$ of $\phi$). To do it, we consider the GR plus scalar field equations
\begin{equation}\label{eq:GmnGR}
{G^\mu}_\nu(q)={\kappa^2}\left[q^{\mu\alpha}\phi_{,\alpha}^{\ast} \phi_{,\nu}-\frac{\mathcal{L}_\phi}{2}{\delta^\mu}_\nu\right] \ ,
\end{equation}
where $\mathcal{L}_\phi=X_\phi+2\,U(\phi,\phi^{\ast})$,  $X_\phi \equiv q^{\alpha\beta}\partial_\alpha\phi^{\ast}  \partial_\beta \phi$, and $\phi_{,\nu}\equiv \partial_\nu \phi$, and by direct comparison between Eqs.(\ref{eq:GmnGR}) and (\ref{eq:GmnScalar}) we establish the following correspondences
\begin{eqnarray}\label{eq:kinetic}
\frac{g^{\mu\alpha}\Phi^*_{,\alpha} \Phi_{,\nu}}{|\hat{\Omega}|^{1/2}}&=& q^{\mu\alpha}\phi^*_{,\alpha} \phi_{,\nu} \ ,  \\
\frac{\mathcal{L}_G-V(\Phi,\Phi^*)}{|\hat{\Omega}|^{1/2}}&=& \frac{\mathcal{L}_\phi}{2} \ . \label{eq:potential}
\end{eqnarray}
Tracing over Eq.(\ref{eq:kinetic}), we get
\begin{equation}\label{eq:mapsca1}
X_\phi=\vert \hat{\Omega} \vert^{-1/2} X_\Phi   \ ,
\end{equation}
and given that $|\hat{\Omega}|^{1/2}$ is a function of $X_\Phi$ and $V(\Phi,\Phi^{\ast})$, this relation implies that $X_{\phi}=F_1[X_\Phi, V]$, {\it i.e.}, it is some (nonlinear) function of the scalar quantities $X_\Phi$ and $V$. Using this result in Eq.(\ref{eq:potential}), elementary algebra yields
\begin{equation} \label{eq:mapsca2}
U(\phi,\phi^{\ast})= |\hat{\Omega}|^{-1/2}\left(\mathcal{L}_G-\tfrac{1}{2}\mathcal{L}_\Phi\right) \ ,
\end{equation}
which implies that $U=F_2[X_\Phi, V]$. This result puts forward that the scalar functions $X_\Phi$ and $V$ can be determined in terms of $X_\phi$ and $U$ by inverting the above relations.

The relevance of this result is now clear: with $X_\Phi$ and $V$ expressed in terms of $X_\phi$ and $U$ for a given solution of GR, and using the relation (\ref{eq:kinetic}) to express ${X^\mu}_\nu$ in terms of $q^{\mu\alpha}\phi^*_{,\alpha} \phi_{,\nu}$, the matrix ${\Omega^\mu}_\nu$ in (\ref{eq:scalar}) is completely determined, which allows one to find the spacetime metric $g_{\mu\nu}$ of the particular RBG considered by means of $q_{\mu\nu}=g_{\mu\alpha}{\Omega^\alpha}_\nu$, with $q_{\mu\nu}$ the solution of GR that follows from solving (\ref{eq:GmnGR}). Remarkably, the algorithm that maps the solutions of GR into solutions of the chosen RBG is independent of the symmetries of the particular solution considered. Therefore, any known solution of GR involving a scalar field, can be mapped into the above family of RBGs. An example will be given later.

\section{Gravity with a fluid}

Our previous analysis may be extended beyond the scalar field case to other sources such as fluids, which are common in the modeling of astrophysical and cosmological setups . For the sake of generality, let us consider an anisotropic fluid with energy-momentum tensor of the form \cite{Herrera97, Gimeno17}
\begin{equation}\label{eq:fluid}
{T^\mu}_\nu=p_{\perp} {\delta^\mu}_{\nu} +(\rho + p_{\perp}) u^{\mu}u_{\nu}+ (p_r-p_{\perp}) \chi^{\mu}\chi_{\nu} \ ,
\end{equation}
where $u^\mu$ and $\chi^{\mu}$ are normalized (with respect to $g_{\mu\nu}$) timelike and spacelike vectors, respectively, subject to the condition $u^{\mu}\chi^{\nu}g_{\mu\nu}=0$, while $\{\rho,p_r,p_{\perp}\}$ represent the energy density, and the radial and tangential pressures of the fluid, respectively.

Similarly as in the scalar field case, any nonlinear function of this ${T^\mu}_\nu$ will possess an identical algebraic structure, thus allowing us to write
\begin{equation}\label{eq:OmFluid}
{\Omega^\mu}_{\nu}=D_1{\delta^\mu}_{\nu} +D_2 u^{\mu}u_{\nu}+ D_3\chi^{\mu}\chi_{\nu} \ ,
\end{equation}
with the $D_i$ being functions of $\rho$, $p_r,$ and $p_{\perp}$, and the indices in  $u_{\nu}$ and $\chi_{\nu}$ are lowered with $g_{\mu\nu}$. Naturally, $\mathcal{L}_G$ and $|\hat\Omega|$ will be functions of those variables too.
Inserting this ${T^\mu}_\nu$ into Eq.(\ref{eq:GmnGeneral}), we get
\begin{eqnarray}\label{eq:BIfluid}
{G^\mu}_{\nu}(q)&=&\frac{\kappa^2}{\vert \hat{\Omega} \vert^{1/2}} \Big[\Big(\frac{\rho-p_r}{2} -\mathcal{L}_G\Big) {\delta^\mu}_{\nu} \nonumber \\
&+& (\rho+p_{\perp}) u^{\mu}u_{\nu} + (p_r-p_{\perp}) \chi^{\mu}\chi_{\nu} \Big] \ .
\end{eqnarray}
Now we parallel the scalar field case and propose a new fluid coupled to GR to explore the possible correspondences among variables.
Introducing similar orthogonal (with respect to $q_{\mu\nu}$) timelike and spacelike vectors in GR, $v^\mu \xi^{\nu}q_{\mu\nu}=0$, and energy density, radial and tangential pressures, denoted by $\{\rho^q,p^q_r,p^q_{\perp}\}$, the corresponding Einstein equations read
\begin{equation}\label{eq:GRfluid}
{G^\mu}_{\nu}(q)=\kappa^2 \left[p^q_{\perp} {\delta^\mu}_{\nu} + (\rho^q+p^q_{\perp}) v^{\mu}v_{\nu} +(p^q_r-p^q_{\perp}) \xi^{\mu} \xi_{\nu} \right] \ ,
\end{equation}
where the indices in  $v_{\nu}$ and $\xi_{\nu}$ are lowered with $q_{\mu\nu}$.
Matching Eqs.(\ref{eq:BIfluid}) and (\ref{eq:GRfluid}) one finds the following correspondences:  $u^{\mu}u_{\nu}=v^{\mu}v_{\nu}$, $ \chi^{\mu}\chi_{\nu} =\xi^{\mu} \xi_{\nu} $, and
\begin{eqnarray}\nonumber
p^q_{\perp}&=&\frac{1}{\vert \hat{\Omega} \vert^{1/2}}[\tfrac{\rho-p_r}{2}-\mathcal{L}_G] \\
\rho^q+p^q_{\perp}&=&\frac{\rho+p_{\perp}}{\vert \hat{\Omega} \vert^{1/2}} \label{eq:BIfluid1} \\
p^q_r-p^q_{\perp}&=&\frac{p_r-p_{\perp}}{\vert \hat{\Omega} \vert^{1/2}} \ . \nonumber
\end{eqnarray}
These last three equations are, in principle, enough to express the three scalars $\{\rho,p_r,p_{\perp}\}$ as functions of $\{\rho^q,p^q_r,p^q_{\perp}\}$. This, along with the vectorial relations $u^{\mu}u_{\nu}=v^{\mu}v_{\nu}$ and $ \chi^{\mu}\chi_{\nu} =\xi^{\mu} \xi_{\nu}$, allows us to express ${\Omega^\mu}_\nu$ in (\ref{eq:OmFluid}) in terms of the solution provided by GR. The metric $g_{\mu\nu}$ is then obtained via $g_{\mu\nu}=q_{\mu\alpha}{(\Omega^{-1})^\alpha}_\nu$.

\section{Examples}

For concreteness, and in order to make contact with recent literature, we will consider the so-called Eddington-inspired Born-Infeld (EiBI) theory of gravity, whose Lagrangian has the form $\LL_G= \frac{1}{ \e \k^2}\left[\vert {\delta^\alpha}_\beta+\epsilon {M^\alpha}_\beta\vert^{1/2}-\lambda\right]$, where $\lambda$ is a constant (very close to unity), $\epsilon$ is a parameter with dimensions of length squared, and ${M^\alpha}_\beta\equiv g^{\alpha\kappa}R_{(\kappa\beta)}(\Gamma)$. This theory recovers GR+$\Lambda_{eff}$ in the limit $\epsilon\to 0$ (with $\Lambda_{eff}={(\lambda-1)/\epsilon}$), and produces deviations from GR at high energy densities for any nonzero value of $\epsilon$ (see \cite{BHOR} for a recent review on Born-Infeld inspired theories).
In this theory, the matrix ${\Omega^\mu}_{\nu}$ satisfies the relation
\begin{equation} \label{eq:veromega}
|\hat{\W}|^{1/2}{(\W^{-1})^\m}_\n=\l {\d^{\m}}_{\n}  -\e\k^2{T^\m}_\n \ ,
\end{equation}
and $\mathcal{L}_G$ takes the on-shell form $\mathcal{L}_G=(|\hat{\W}|^{1/2}-\lambda)/\epsilon \kappa^2$.
For the scalar field case this algebraic equation leads to
\begin{equation}\label{eq:OmscalarBI}
{\Omega^\mu}_{\nu}=\frac{\tilde{\lambda}^{1/2}}{(\tilde{\lambda}-2\tilde{X}_{\Phi})^{1/2}}\left[(\tilde{\lambda}-2\tilde{X}_{\Phi}) {\delta^\mu}_{\nu} +2{\tilde X}^\mu_{\ \nu}  \right] \ ,
\end{equation}
and $\vert \hat{\Omega} \vert^{1/2}=\tilde{\lambda}^{3/2}(\tilde{\lambda}-2\tilde{X}_{\Phi})^{1/2} $, where $\tilde{\lambda}=\lambda + \epsilon \kappa^2 V+\tilde{X}_{\Phi}$, with ${\tilde X}^\mu_{\ \nu} \equiv (\epsilon \kappa^2/2) g^{\mu\alpha} \Phi^*_{,\alpha}\Phi_{,\nu}$, and $\tilde{X}_{\Phi} \equiv(\epsilon \kappa^2/2) X_{\Phi}$.

The coupling of this theory to a \emph{free massless} real scalar field was studied in \cite{AOR17}, where numerical methods were used to solve the field equations (with $\lambda=1$). The same system was considered in GR by Wyman \cite{Wyman}, obtaining analytical solutions. Now we can use the results presented above to find an analytical solution for the EiBI theory which can be compared with the numerical results of \cite{AOR17}. Focusing on asymptotically flat, static, spherically symmetric configurations, the GR solution can be suitably cast in terms of the line element (in units such that $\kappa^2=1$, and using a subindex $q$ to denote the GR solution)
\begin{equation} \label{eq:lineelGR}
d{s}_{q}^2=-e^{{\nu_q}}dt^2+ \frac{1}{{W_q}^2}\left(d\theta^2+\sin^2\theta d\varphi^2\right)+\frac{dy^2}{ {W_q}^4e^{-{\nu}_q} } \ ,
\end{equation}
where
\begin{equation} \label{eq:Wymanfun}
{\nu_q}=\alpha y \ ;\q {W_q}= e^{\alpha y/2}(\sinh (\gamma y)/\gamma) \ ,
\end{equation}
with $\alpha$ an integration constant and $\gamma\equiv\sqrt{\alpha^2+2}/2$. In the far limit, which corresponds to $y\to 0$, one must take $\alpha=-2M$ in order to recover the Newtonian limit. In the above line element $ \phi=y$, {\it i.e.}, the scalar field is being used as the radial coordinate, see \cite{AOR17,Wyman}. Now, combining Eqs. (\ref{eq:mapsca1}) and (\ref{eq:mapsca2}) in the $V=U=0$ case, one obtains
\begin{equation}
\vert \hat{\Omega} \vert^{1/2}=\frac{\lambda}{1-\tilde{X}_{\phi}} \ ,
\end{equation}
where $\tilde{X}_{\phi}\equiv(\epsilon \kappa^2/2) X_{\phi} = (\epsilon \kappa^2/2) {W_q}^4 e^{-{\nu_q}}$. Replacing this expression back into Eq.(\ref{eq:mapsca1}) one finds
\begin{equation} \label{eq:scalargeon}
\tilde{X}_{\Phi}=\frac{\lambda \tilde{X}_{\phi}}{1-\tilde{X}_{\phi}} \ .
\end{equation}
Particularizing (\ref{eq:OmscalarBI}) to this example, one obtains ${\Omega^\alpha}_{\beta}={\rm diag}(\Omega_{+},\Omega_{+},\Omega_{+},\Omega_{-})$, with $\Omega_{+}=(\lambda^2-\tilde{X}_{\Phi}^2)^{1/2}$ and $\Omega_{-}=(\lambda+\tilde{X}_{\Phi})^{3/2}(\lambda-\tilde{X}_{\Phi})^{-1/2}$. Defining now the line element of the spacetime metric of EiBI theory as in (\ref{eq:lineelGR}) but omitting the subindex $q$, the metric functions take the form \cite{AOR17}
\begin{equation} \label{eq:geonrel}
e^{\nu}=e^{{\nu_q}}/\Omega_{+} \,;\q W^2=\Omega_{+}{W^2_q}  \ .
\end{equation}
In \cite{AOR17}, solutions were obtained by considering Eqs.(\ref{eq:GmnScalar}), which are highly nonlinear in the ${T^{\mu}}_\nu$ of $\Phi$,
specifying a value for the constant $\alpha$ in the far region ($y\to 0$), where the GR solution is recovered, and numerically integrating towards $y\to \infty$, where approximate analytical solutions are also available. The numerics allowed us to determine the value of the coefficients of the approximate far solutions. Using instead Eqs. (\ref{eq:geonrel}), which are linear in the ${T^{\mu}}_\nu$ of $\phi$, one finds analytical expressions for those coefficients in terms of $\alpha$, exactly reproducing the results of the numerical integration. In particular, the coefficient $l_2$ in  Table II and Fig. 3 of Ref.\cite{AOR17}, is obtained by solving $l_2+\sqrt{l_2^2+4}/2=\alpha+\sqrt{\alpha^2+2}/2$ as $l_2=l_2(\alpha)$. Similar expressions exist for the other parameters in that table. This nicely illustrates the power of our method, which allows us to obtain analytical solutions for a highly nonlinear self-gravitating scalar field $\Phi$ in the Einstein frame of the theory in terms of a linear field $\phi$ in that same frame. Interestingly, this scalar field solution exhibits a wormhole structure above a certain low mass threshold, with the throat located at $r = 2M$, while in GR the topology is simple.

Let us now consider the anisotropic fluid in the EiBI theory. With a bit of  algebra, Eqs.(\ref{eq:BIfluid1}) yield
\begin{eqnarray}
\lambda+\tilde{\rho}&=&\sqrt{\frac{1+ \left[\tilde{p}^q_{\perp}+\tfrac{(\tilde{\rho}^q+\tilde{p}^q_r)}{2}\right]}{1+ \left[\tilde{p}^q_{\perp} -\tfrac{(\tilde{\rho}^q+\tilde{p}^q_r)}{2} \right]}} \frac{1}{[1+\frac{(\tilde{p}^q_r-\tilde{\rho}^q)}{2}]} \label{eq:fluid1}  \\
\lambda-\tilde{p}_r&=&\sqrt{\frac{1+ \left[\tilde{p}^q_{\perp}-\tfrac{(\tilde{\rho}^q+\tilde{p}^q_r)}{2}\right]}{1+ \left[\tilde{p}^q_{\perp} + \tfrac{(\tilde{\rho}^q+\tilde{p}^q_r)}{2} \right]}}  \frac{1}{[1+\frac{(\tilde{p}^q_r-\tilde{\rho}^q)}{2}]}  \label{eq:fluid2}  \\
\lambda-\tilde{p}_{\perp}&=&  \frac{1}{\sqrt{1+\left[ \tilde{p}^q_{\perp} +\frac{(\tilde{\rho}^q+\tilde{p}^q_r)}{2} \right]}\sqrt{1+\left[ \tilde{p}^q_{\perp} -\frac{(\tilde{\rho}^q+\tilde{p}^q_r)}{2} \right]}}  \ , \label{eq:fluid3}
\end{eqnarray}
where the tildes indicate that a factor $\epsilon \kappa^2$ is implicit, {\it i.e.}, $\tilde{\rho} \equiv \epsilon \kappa^2 \rho$, and so on. Since the coefficients $D_1$, $D_2,$ and $D_3$ of ${\Omega^\mu}_\nu$ in (\ref{eq:OmFluid}) depend on $\rho$, $p_r,$ and $p_\perp$ (explicit expressions can be easily obtained), once a solution of the GR problem is given, ${\Omega^\mu}_\nu$ and $g_{\mu\nu}$ can be explicitly computed.

A simple example of this class of anisotropic fluids is given by models of nonlinear electrodynamics (NED), which satisfy $p^q_r=-\rho^q$ and $p^q_{\perp}=K(\rho^q)$, where the function $K(\rho^q)$ specifies such a model, with $K(\rho^q)=\rho^q$ Maxwell's theory. Interestingly, a NED model in the Einstein frame leads to another NED in the RBG frame, as can be easily verified from Eqs.(\ref{eq:fluid1}), (\ref{eq:fluid2}), and (\ref{eq:fluid3}), which yield
\begin{equation}
\tilde{p}_r=-\tilde{\rho}=\frac{-\lambda \tilde{\rho}^q + (\lambda-1)}{1-\tilde{\rho}^q} \ ; \q
\tilde{p}_{\perp}=\frac{\lambda  \tilde{p}^q_{\perp}+(\lambda-1)}{1+\tilde{p}^q_{\perp}} \ .
\end{equation}
If one is interested in coupling the EiBI theory to Maxwell's electrodynamics, for which $\tilde{p}_{\perp}=\tilde\rho$, then on the GR side one finds a NED of the form (for $\lambda=1$) $\tilde p^q_{\perp}=\tilde\rho^q/(1-2\tilde{\rho}^q)$. Solving the GR equations for that NED, one can recover the EiBI+Maxwell solution previously obtained in \cite{Banados:2010ix,EMor} using the original frame. The validity of the method has also been verified with other RBGs coupled to different NEDs  \cite{bor17a,MOR17}. In the isotropic case, $\tilde{p}_r=\tilde{p}_\perp$, one recovers the results of \cite{Delsate:2012ky}. It is worth noting that these charged solutions in metric-affine EiBI and $f(R)$ theories typically give rise to wormhole structures. This signals a rather generic presence of such objects beyond GR and may have nontrivial implications in dark matter scenarios  \cite{Lobo:2013prg} and for the stability of primordial black holes \cite{Olmo:2013mla}.

\section{Conclusion}

In this work we have shown that for any metric-affine RBG coupled to scalar fields or anisotropic fluids the solutions can be generated out of an analogous problem in GR. This is possible thanks to a correspondence worked out in the Einstein frame that turns the right-hand side of (\ref{eq:GmnGeneral}), which is a highly nonlinear function of the original ${T^\mu}_\nu$, into the ${T^\mu}_\nu$ of an analogous field, in this way transforming the original modified gravity problem into a standard problem in GR.
The map that relates the solutions only depends on the particular kind of matter source and is independent of the symmetries of the specific configuration considered, thus establishing a correspondence between the whole spaces of solutions. The method should, in principle, also be implementable in many other scenarios. For vector fields, for instance, one expects that the algebraic structure of the matrix ${\Omega^\mu}_\nu$ will typically require more terms than those present in ${T^\mu}_\nu$. The case of fermions requires further investigation due to the nontrivial/nonuniversal role of torsion in those cases \cite{Afonso:2017bxr}. Scenarios with several sources are also possible and will lead to mixing between different fields in ${\Omega^\mu}_\nu$, adding technical difficulties to the analysis.

The results presented here can be used to work out static, stationary, and fully dynamical configurations in RBGs. Extensions of the Kerr-Newman solution, rotating hairy black holes, mergers of compact objects, perturbations, cosmological scenarios, higher-dimensional braneworld models, and any other solution of physical or mathematical interest \cite{Stephani:2003ika} can now be implemented in a large family of gravity theories taking advantage of the analytical and numerical methods and techniques developed for GR. The robustness of some predictions and the confrontation of these theories with observations can now be tackled in a convenient and systematic way, opening new avenues to explore new gravitational physics beyond GR.

\section{Acknowledgements}

This work was supported by the Ramon y Cajal Contract No. RYC-2013-13019 (Spain), the FCT (Portugal) Grants No. SFRH/BPD/102958/2014 and No. UID/FIS/04434/2013, the Spanish Project No. FIS2014-57387-C3-1-P (MINECO/FEDER, EU), the Project No. H2020-MSCA-RISE-2017 (FunFiCO-777740), Project No. SEJI/2017/042 (Generalitat Valenciana), the Consolider Program CPANPHY-1205388, and the Severo Ochoa Grant No. SEV-2014-0398 (Spain). This article is based upon work from COST Action CA15117, supported by COST (European Cooperation in Science and Technology).

\end{document}